\documentclass[twocolumn,prl,showpacs,amsmath,amssymb]{revtex4}
\usepackage{graphicx,amsmath,epsfig,latexsym,color}
\usepackage{comment}
\newcommand{\be}{\begin{equation}}
\newcommand{\ee}{\end{equation}}
\begin{document}
\title{Coherent Atom-Molecule Oscillations in a Bose-Fermi Mixture}
\author{M. L.\ Olsen}
\author{J. D.\ Perreault}
\email{johndp@jilau1.colorado.edu}
\author{T. D.\ Cumby}
\author{D. S.\ Jin}
\affiliation{JILA, National Institute of Standards and Technology
and University of Colorado, and Department of Physics, University of
Colorado, Boulder, Colorado  80309-0440, USA}
\date{\today}
\begin{abstract}

We create atom-molecule superpositions in a Bose-Fermi mixture of
$^{87}$Rb and $^{40}$K atoms. The superpositions are generated by
ramping an applied magnetic field near an interspecies Fano-Feshbach
resonance to coherently couple atom and molecule states. Rabi- and
Ramsey-type experiments show oscillations in the molecule population
that persist as long as 150 $\mu$s and have up to 50\% contrast. The
frequencies of these oscillations are magnetic-field dependent and
consistent with the predicted molecule binding energy.  This quantum
superposition involves a molecule and a pair of free particles with
different statistics (i.e. bosons and fermions), and furthers
exploration of atom-molecule coherence in systems without a
Bose-Einstein condensate.
\end{abstract}
\pacs{34.80.Pa, 05.30.Jp, 05.30.Fk, 34.50.-s} \keywords{Feshbach
resonance, ultracold molecules} \maketitle

Magnetic-field tunable Fano-Feshbach resonances in ultracold atom
gases provide a quantum system that can couple atoms to a molecular
bound state \cite{kohl06}. This coupling has been used to associate
ultracold atoms into weakly bound molecules by adiabatically
sweeping a magnetic field across a Fano-Feshbach resonance
\cite{regal03}.  Quantum superpositions of molecules and
\emph{single} atom-pair states have also been created with these
resonances, starting from a gas of bosonic atoms
\cite{donl02,thom05,syas07,winkler05,Ryu2005}. These atom-molecule
superpositions are highly unusual in that they establish coherence
between the chemically distinct states of a free atom pair and a
bound molecule.  In experiments using identical bosons either in a
Bose-Einstein condensate (BEC) \cite{donl02} or confined pairs in a deep
optical lattice potential \cite{syas07}, the atom-molecule
oscillations are well described by a two-level model.  However,
observations of atom-molecule superpositions in thermal gases of
bosons \cite{thom05,dumk05} extend coherent atom-molecule
manipulation to more generic systems where the coherent
atom-molecule oscillations involve a \emph{continuum} of atom-pair
states with different relative momenta.  This raises the possibility
of observing atom-molecule oscillations in a many-body system
involving fermions, which is a subject that has been investigated
theoretically
\cite{Dannenberg2003,Wouters2003,Andreev2004,Barankov2004,Stecher2007}.
In this Rapid Communication, we demonstrate atom-molecule superpositions in a
Bose-Fermi mixture of $^{87}$Rb and $^{40}$K, where we use
magnetic-field pulses near a Fano-Feshbach resonance to reveal
coherent oscillations that persist for as long as \mbox{150 $\mu$s}. Our
work furthers exploration of atom-molecule coherence in non-BEC
systems, and provides the groundwork for possible future
experiments.  For example, the frequency, coherence time, and amplitude of such
atom-molecule oscillations can be used to probe non-equilibrium
dynamics and pairing processes in strongly interacting Bose-Fermi
\cite{Wouters2003} and Fermi-Fermi
\cite{Andreev2004,Barankov2004,Stecher2007} atom gases.
Understanding atom-molecule coupling in this intrinsically
multi-level system is also relevant to the goal of efficient
molecule creation in trapped ultracold atom gases \cite{ni08}.

Efficient molecule creation using a Fano-Feshbach resonance requires
an initial atom gas with high phase space density
\cite{hodb05,kohl06}. We prepare an ultracold Bose-Fermi mixture of
$^{87}$Rb atoms in the $|f,m_{f}\rangle$ = $|1, 1\rangle$ state and
$^{40}$K atoms in the $|9/2, -9/2\rangle$ state
\cite{gold04,inou04}. Here, $f$ is the total atomic spin, and
$m_{f}$ is the spin projection along the magnetic field. The atoms
are confined in a far-off-resonance optical dipole trap formed by a
single laser beam with a waist of 18 $\mu$m and a wavelength of \mbox{1090 nm}.
The gas mixture is evaporatively cooled by decreasing the power
of the optical-trap beam. Following the evaporation, the optical
trap power is adiabatically increased, and the measured trap
frequencies are 350 Hz for Rb and 490 Hz for K in the radial
direction and 5.2 Hz for Rb and 8.1 Hz for K in the axial direction.
In this trap, we have $N_{\mathrm{Rb}}=8\times10^{4}$ Rb atoms and
$N_{\mathrm{K}}=2\times10^{5}$ K atoms at a temperature of 200 nK,
which corresponds to 1.2 $T_{C}$ and 0.3 $T_{F}$, where $T_{C}$ is
the critical temperature for Bose-Einstein condensation of the Rb
gas and $T_{F}$ is the Fermi temperature of the K gas.

Molecule detection is achieved with spin-selective absorption
imaging, using light tuned to the K $4S_{1/2}|9/2,
-9/2\rangle\rightarrow4P_{3/2}|11/2, -11/2\rangle$ cycling
transition \cite{Ospelkaus2006,Zirbel2008b}.  To avoid imaging
unbound K atoms, we use a 30 $\mu$s pulse of rf tuned to the K
Zeeman transition to transfer the atoms to the $|9/2,-7/2\rangle$
state with 99\% efficiency. The molecules are unaffected since their
binding energy of $h\times$140 kHz (at 546.04 G) is larger than the
spectral width of the rf pulse. The molecule cloud is imaged after 2
ms of expansion from the optical trap.

To characterize the atom-molecule coupling time scale, we first
explored molecule creation via linear magnetic-field sweeps across
the Fano-Feshbach resonance between Rb $|1,1\rangle$ and K
$|9/2,-9/2\rangle$ atoms at $B_{0}=546.76$ G \cite{inou04}. The
atoms are initially prepared by turning on an applied magnetic field
of \mbox{542 G} during the evaporation.  The magnetic field is then
increased to the high-field side of the resonance (547.2 G) after
evaporation, at a speed of $\dot{B}_{\rm{fast}}=140$ G/ms.  Next, we
sweep the magnetic field down through the resonance to 546.04 G and
measure the number of molecules created by the sweep. The measured
number of molecules created as a function of the magnetic-field ramp
rate $\dot{B}$ fits well to an exponential function
$N_{\rm{mol}}=N_{\rm{max}}(1-e^{-\beta/\dot{B}})$, where the two fit
parameters are the saturated number of molecules for the slowest
ramps \mbox{$N_{\rm{max}}=(2.9\pm0.3)\times10^{4}$} and $1/e$ ramp
speed $\beta=(26\pm8)$ G/ms. For these data, the calculated peak
number densities of the initial atom gas are $n_{\rm{Rb}}^{0} =
1.1\times10^{13}$ cm$^{-3}$ and $n^{0}_{\rm{K}}=1.3\times10^{13}$
cm$^{-3}$, for Rb and K atoms, respectively. The saturated molecule
number $N_{\rm{max}}$ is 36\% of $N_{\rm{Rb}}$, which as the smaller
atom number sets an upper limit on the number of molecules.
$N_{\rm{max}}$ is significantly lower than the prediction of a
phenomenological model that has been shown to successfully predict
molecule conversion \cite{hodb05,papp06,Zirbel2008b}.  This may be
due to collisional losses or heating near the resonance.

\begin{figure}
\scalebox{.8}{\includegraphics{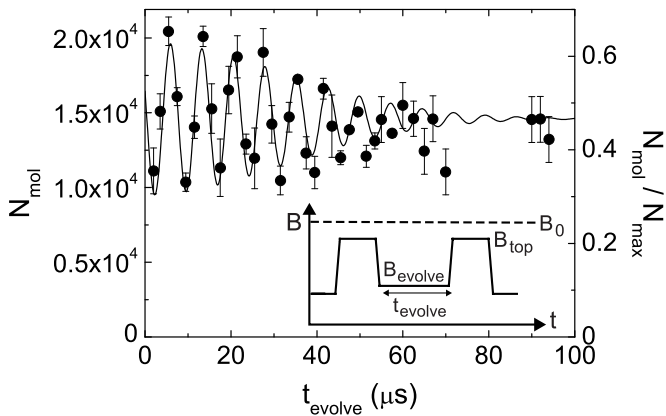}}
\caption{\label{fig:RamseyOsc}Ramsey-type atom-molecule
oscillations. The molecule number $N_{\rm{mol}}$ is shown as a
function of the hold time at $B_{\rm{evolve}}=546.08$ G. Here, we
fit the data to $N_{\rm{mol}}=A
e^{-t^{2}/(2\sigma^{2})}$sin$(2\pi\nu_{\rm{osc}}t-\phi)+y_{0}$,
which would describe, for example, oscillations that dephase due to
inhomogeneous broadening. The fit gives an oscillation frequency
$\nu_{\rm{osc}}=(136.5\pm1.3)$ kHz and rms damping time
$\sigma=(32\pm5)$ $\mu$s. The data fit equally well to an
exponentially damped oscillation with a 1/$e$ damping time of
$(43\pm12)$ $\mu$s. (Inset) Schematic of magnetic-field pulse
sequence.}
\end{figure}

To create atom-molecule superpositions we use non-adiabatic ramps
with speeds that are fast compared to $\beta$, similar in technique
to references \cite{donl02,syas07}. We first consider atom-molecule
coherence far from the resonance, where we use a double-pulse
experiment that is analogous to Ramsey's method of separated
oscillatory fields \cite{ramsey50,Goral2005,kohl06}. A schematic of
the magnetic-field ramps is shown in the inset of Fig.\
\ref{fig:RamseyOsc}. The two pulses toward resonance couple the atom
and molecule states. The duration and magnetic field of the pulses
are empirically optimized for maximum amplitude oscillations in the
molecule population.  The pulse sequence begins after a slow (3
G/ms) ramp to 545.80 G and consists of two trapezoidal pulses with
15 $\mu$s holds at $B_{\rm{top}}=546.58$ G, separated by a variable
hold time $t_{\rm{evolve}}$ at $B_{\rm{evolve}}$. The outer ramps
have speeds of $\dot{B}_{\rm{fast}}$, while the inner ramps have
speeds of $(B_{\rm{top}}-B_{\rm{evolve}})/5 \mu$s \cite{fastB}. At
$B_{\rm{top}}$, the calculated molecule size is 1\,800 $a_{0}$,
which is comparable to the typical distance between nearest-neighbor
Rb and K atoms, which is approximately $n^{-1/3} = 10\,500 a_{0}$.
Here, $n =\frac{1}{N_{<}}\int n_{\rm{K}}(r)n_{\rm{Rb}}(r)d^{3}r$,
where $n_{\rm{K}}$ and $n_{\rm{Rb}}$ are the number densities of K and
Rb, respectively, and $N_{<}$ is the number of atoms in the species
with fewer atoms.

Figure \ref{fig:RamseyOsc} shows the measured molecule number after
a double-pulse sequence with $B_{\rm{evolve}}=546.08$ G. The
right-hand axis shows the molecule number $N_{\rm{mol}}$ normalized
by $N_{\rm{max}}$. The molecule number shows clear oscillations as a
function of $t_{\rm{evolve}}$, as expected for a coherent
atom-molecule superposition. We observe a peak-to-peak amplitude of
$10^{4}$ molecules, which is 13\% of $N_{\rm{Rb}}$.

\begin{figure}
\scalebox{.85}{\includegraphics{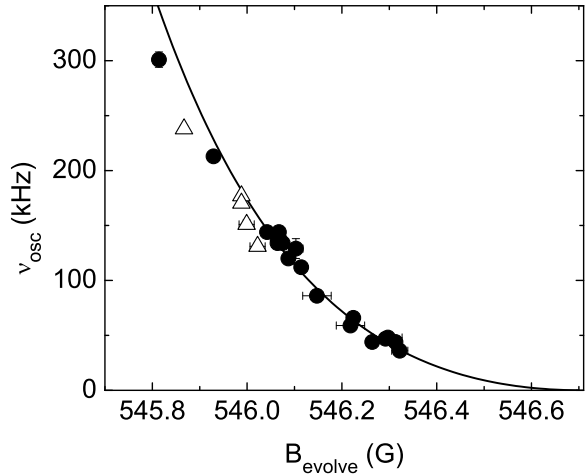}} \caption{\label{fig:bindingenergy}Magnetic-field dependence of the
measured molecule number oscillation frequency in Ramsey-type double-pulse experiments. The circles ($\bullet$)
correspond to double-pulse experiments performed with the atoms confined in the optical trap, while the triangles
($\triangle$) represent experiments performed after 1 ms of expansion from the trap. The solid line is a fit to
the universal prediction for molecule binding energy \cite{Gribakin1993} for the data above 546 G \cite{freqfit}.}
\end{figure}

As further evidence that these oscillations are due to atom-molecule superpositions, we have measured the
oscillation frequency for various values of $B_{\rm{evolve}}$ and find that the frequency corresponds to the
predicted binding energy of the molecules. Figure \ref{fig:bindingenergy} shows the measured oscillation frequency
$\nu_{\rm{osc}}$ as a function of the hold magnetic field $B_{\rm{evolve}}$. The solid curve is a fit to the
universal prediction for the molecule binding energy near the resonance,
$E_{b}=\hbar^{2}/[2\mu_{\rm{KRb}}(a-\bar{a})^2]$ \cite{Gribakin1993}. Here, $\mu_{\rm{KRb}}$ is the $^{40}$K and
$^{87}$Rb reduced mass, $\bar{a} = 68.8 a_{0}$, and $a=a_{\rm{bg}}[1-\Delta/(B-B_{0})]$ with $a_{\rm{bg}}=-185
a_{0}$ \cite{Ferlaino2006e}. From the fit, we extract the resonance position $B_{0}=(546.71\pm0.01)$ G and width
$\Delta=(-3.34\pm0.05)$ G, in agreement with Ref.\ \cite{Zirbel2008b}.

Figure \ref{fig:bindingenergy} also includes data taken at
significantly lower atom number densities.  Here, we lower the
density of the atoms by turning off the optical trap and allowing
the gas to expand before applying the magnetic-field pulses. For
these data, at the end of evaporation, we have $1\times10^{6}$ Rb
atoms and $6\times10^{5}$ K atoms at 1200 nK in an optical trap with
radial trapping frequencies of 690 and 970 Hz for Rb and K,
respectively. The temperature corresponds to $T/T_{C}=1.6$ for Rb
and $T/T_{F}=0.7$ for K.  The triangles in Fig.\
\ref{fig:bindingenergy} correspond to experiments performed after \mbox{1 ms}
of expansion using a hold time of 50 $\mu$s at the top of the
optimized pulses. After 1 ms of expansion, the peak densities are
$n^{0}_{\rm{Rb}}=3\times10^{12}$ cm$^{-3}$ and
$n^{0}_{\rm{K}}=7\times10^{11}$ cm$^{-3}$. By performing linear
magnetic-field sweeps as discussed earlier, we find
$N_{\rm{max}}=(9.2\pm0.6)\times10^{4}$, which is 15\% of
$N_{\rm{K}}$, and $\beta=(9.1\pm1.0)$ G/ms.

\begin{figure}
\scalebox{.85}{\includegraphics{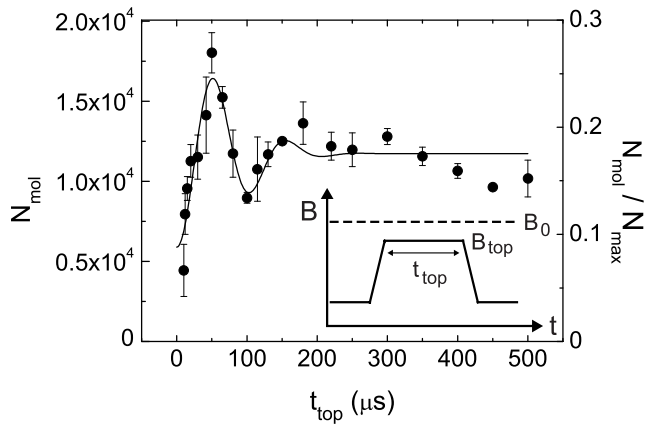}}
\caption{\label{fig:RabiOsc}Rabi-type atom-molecule oscillations.
The measured molecule number is shown as a function of the hold time
for a single pulse where $B_{\rm{top}}$ was 546.51 G. The line is a
fit to $N_{\rm{mol}}=A
e^{-t^{2}/(2\sigma^{2})}$cos$(2\pi\nu_{\rm{osc}}t)+y_{0}$, yielding
a frequency $\nu_{\rm{osc}}=(9.3\pm0.5)$ kHz and rms damping time
\mbox{$\sigma=(80\pm20)$ $\mu$s}. If we fit the data to an
exponentially damped oscillation, the 1/$e$ damping time is
$(80\pm30)$ $\mu$s. (Inset) Schematic of the magnetic-field pulse.}
\end{figure}

For values of $B_{\rm{evolve}}$ that approach the resonance, the
double magnetic-field pulse shown in the inset to \mbox{Fig.
\ref{fig:RamseyOsc}} would begin to resemble a single pulse.
Therefore, to study atom-molecule oscillations at magnetic fields
closer to the resonance, we employ single magnetic-field pulses that
realize a Rabi-type experiment. Previously, oscillations were seen
in single-pulse experiments with a single species of bosons
\cite{donl02,thom05,syas07}. However, in the BEC system,
oscillations in these Rabi-type experiments proved more difficult to
observe than those in double-pulse experiments \cite{donl02}. In our
case, we find that we are only able to observe oscillations in a
single-pulse experiment at the lower densities enabled by expansion
of the gas.  For the higher density trapped gas, the coherent
atom-molecule oscillations seen in the double-pulse experiments
prove that there must be coherent evolution during each single
pulse, however at these densities it appears that the coherence time
at $B_{\rm{top}}$ is too short to permit the observation of even one
full Rabi oscillation.

A schematic of a single pulse is shown in the inset of Fig.\
\ref{fig:RabiOsc}. The magnetic field is ramped from 545.80 G to
$B_{\rm{top}}$, held for a time $t_{\rm{top}}$, and ramped back
down, with both ramps having speeds of $\dot{B}_{\rm{fast}}$. Figure\
\ref{fig:RabiOsc} shows the measured molecule population in the
expanded gas after a single pulse to $B_{\rm{top}}=$ 546.51 G as a
function of the hold time $t_{\rm{top}}$. Defining the contrast as
the $t_{\rm{top}}=0$ amplitude $|A|$ divided by the final level of
the damped oscillation $y_{0}$, we measure a contrast of
$0.5\pm0.2$.

\begin{figure}
\scalebox{0.9}{\includegraphics{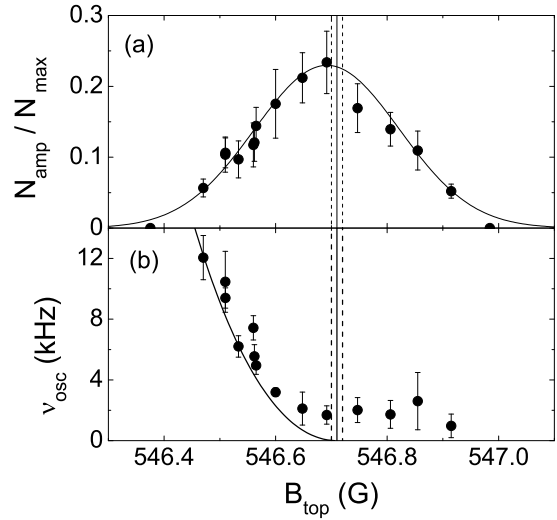}}
\caption{\label{fig:RabiLineshape}Magnetic-field dependence of the
(a) peak-to-peak amplitude and (b) frequency of the single-pulse
oscillations. The solid line in (a) is a Gaussian fit to the data,
while the solid line in (b) is the universal prediction for the
molecule binding energy from Fig. \ref{fig:bindingenergy}. The
vertical lines represent the fitted Fano-Feshbach resonance position
and uncertainty from Fig.\ \ref{fig:bindingenergy}.}
\end{figure}

In optimizing the peak-to-peak amplitude of the oscillations in the
molecule population, we find that the amplitude depends on the value
of $B_{\rm{top}}$ as shown in \mbox{Fig.\ \ref{fig:RabiLineshape}a}. In
addition, as $B_{\rm{top}}$ is increased, we observe fewer
oscillations before they damp out, and for the data above
$B_{\rm{top}}=546.6$ G, only one period of the oscillation is
observed. Therefore, we define the peak-to-peak amplitude as the
difference in molecule number between the first maximum and the
subsequent minimum. The amplitude of the oscillations is peaked near
the Fano-Feshbach resonance where we observe a peak-to-peak
amplitude that is 23\% of $N_{\rm{max}}$. At $B_{\rm{top}}=546.47$
G, the oscillation amplitude drops to 6\% of $N_{\rm{max}}$. Here,
the calculated molecule size is 1\,100 $a_{0}$, which is 5\% of the
typical distance between nearest-neighbor K and Rb atoms of 23\,700
$a_{0}$.

The frequency of the Rabi-type oscillations also depends on
$B_{\rm{top}}$ as shown in Fig.\ \ref{fig:RabiLineshape}b.  Below
the resonance, the measured frequency agrees with the prediction for
the molecule binding energy.  Above 546.6 G, the frequency obtained
by fitting to a Gaussian-damped oscillation saturates, which can be
understood in terms of a \emph{multi-level} Landau-Zener diagram
\cite{Borca2003,juli04}.  This is in contrast to the
\emph{two-level} atom-molecule system explored by Syassen \emph{et
al}.\ \cite{syas07}, in which the measured frequency as a function
of magnetic field is symmetric about the resonance.  In our
experiment, the molecule state can couple to many closely spaced
atom pair energy levels \cite{Borca2003,juli04}.  The energy spacing
of these levels could set a lower limit on the oscillation
frequency. We note that the observed saturation frequency of
$(1.7\pm0.8)$ kHz is on the order of the expected many-body level
spacing of 0.6 kHz \cite{Borca2003}.

The results of the single-pulse experiments can be used to infer the
best pulses for double-pulse experiments.  Maximum contrast should
be achieved by using pulses where the hold time at $B_{\rm{top}}$
gives one quarter cycle of the single-pulse oscillation, which is
analogous to a $\pi/2$ pulse in Ramsey's experiments
\cite{ramsey90}. In fact, we find that the empirically optimized
pulse sequence for the expanded clouds corresponds to this
condition.

For both the single-pulse and double-pulse experiments, we observe
damping of the atom-molecule oscillations. Because the time-averaged
number of molecules does not decrease, this damping is not due to a
loss of molecules. Indeed, we measure molecule lifetimes longer than
a millisecond over the range of densities and fields probed
\cite{Zirbel2008}. The loss of contrast is instead likely caused by
dephasing due to a range of oscillation frequencies among the pairs.
For our Bose-Fermi gas mixture, a range of oscillation frequencies
is expected because there is a spread in the relative kinetic energy
of the pairs, $E_{\mathrm{rel}}$.  The oscillation frequency for a
given atom pair is $\nu_{\mathrm{pair}}=(E_{b}+E_{\mathrm{rel}})/h$.
As a rough estimate of this damping time, we have performed a
semiclassical Monte Carlo calculation of the expansion and find that
the distribution of relative kinetic energies for nearest-neighbor
Rb and K atoms would give a 75 $\mu$s $1/e$ damping time. While this
agrees well with the data shown in Fig.\ \ref{fig:RabiOsc}, we
observe faster damping rates for oscillations farther from the
Fano-Feshbach resonance. This suggests that a technical source of
dephasing, such as a spatial inhomogeneity in the applied magnetic
field, may be important. Given the quadratic dependence of $E_{b}$
on magnetic field, a spatial variation in magnetic field would
result in more rapid dephasing for larger magnetic-field detuning
from the Fano-Feshbach resonance. Future efforts will focus on
minimizing technical causes of dephasing and investigating the
intrinsic dephasing due to the relative kinetic energies of the atom
pairs.

In conclusion, we have observed atom-molecule Rabi- and Ramsey-type
oscillations using a mixture of bosons and fermions.  We do not
start with a BEC or prepare the system in a single energy state
\cite{donl02,syas07}, but rather extend previous work by coherently
coupling atoms with a distribution of energies \cite{thom05,dumk05}
to a fermionic molecule state.  We observe relatively large
amplitude oscillations that persist up to 150 $\mu$s.  The frequency,
coherence time, and amplitude of these atom-molecule oscillations may
provide a unique way to probe the many-body behavior of a strongly
interacting Bose-Fermi mixture \cite{Wouters2003}. Additionally, it
will be interesting to explore decoherence mechanisms for this
non-Bose condensed superposition.

We thank C. H. Greene, E. A. Cornell, and the JILA BEC group for
useful discussions.  We acknowledge funding from NIST and NSF.  \mbox{J.D.P.}
acknowledges support from a NRC Research Associateship Award at
NIST.

%\bibliography{AMosc}

\begin{thebibliography}{30}
\bibitem[{fas()}]{fastB}
 \bibinfo{note}{The maximum magnetic-field ramp speed is limited by the
  bandwidth of the servo. The time-dependent B-field, measured using rf
  spectroscopy on the K atoms, is within 40 mG of the desired field over the
  duration of the pulse.}
\end{thebibliography}

\end{document}